\documentclass[%
 reprint,
superscriptaddress,
 amsmath,amssymb,
 aps,
pra,
]{revtex4-1}

\usepackage{graphicx}% Include figure files
\usepackage{dcolumn}% Align table columns on decimal point
\usepackage{bm}% bold math
\begin{document}

%\preprint{APS/123-QED}

\title{The effect of the geometric potential and an external magnetic field on a charged particle on a helicoid}% Force line breaks with \\
\author{Miguel Antonio Sulangi}
\affiliation{Department of Physics and Astronomy, Johns Hopkins University, Baltimore, MD 21218, USA}
\affiliation{Department of Physics, Ateneo de Manila University, Quezon City, Philippines 1108}
\author{Quirino M. Sugon, Jr.}
\affiliation{Manila Observatory and Department of Physics, Ateneo de Manila University, Quezon City, Philippines 1108 }

\date{November 24, 2012}% It is always \today, today,

\begin{abstract}
We perform an analysis of the combined effects of geometry and a magnetic field for the case of a charged particle on a helicoid. The effective quantum potentials for a charged spinless particle confined on a helicoid for two simple magnetic field configurations are derived. These potentials depend nontrivially on the surface curvature and the external magnetic field. We find that the \emph{qualitative} behavior of the effective potential can be altered by changing the strength of the applied magnetic field. The application of a magnetic field results in effective potentials that are either repulsive or attractive, depending on the magnitude of the magnetic field and the angular momentum of the particle. Finally, for the case of effective potentials that have a minimum, we also obtain approximate expressions for the energy levels valid when the particle is near a minimum, and these are found to be similar in form to the energy levels of a particle in a harmonic oscillator potential. \end{abstract}

\maketitle

%\tableofcontents

\section{\label{sec:level1}Introduction}

Recently, there has been an emergence of methods for experimentally producing nanostructures with novel geometries. A number of studies have demonstrated qualitatively unusual behavior of charge carriers in curved nanostructures in strong external electric and magnetic fields. Physical phenomena that emerge from a combination of strong electric and magnetic fields and surface curvature include Aharonov-Bohm oscillations \cite{aB99}, formation of Landau levels \cite{hA92, jK92}, and the quantum Hall effect \cite{eP07}. Also, graphene systems such as fullerines---which are curved nanostructures---exhibit unusual phenomena induced by strong magnetic fields such as changes in the energy band gap \cite{uC04, bL07} and in the magnetoresistance \cite{nS07, aV07}. 

The helicoid is one surface that has been the subject of intense mathematical study. In particular, the effects of surface curvature on the wavefunction of a particle constrained on a helicoid have already been studied by Atanasov \emph{et al.} \cite{vA09}. However, the problem of a charged particle constrained on the surface of a helicoid in external magnetic fields has not yet been studied. The motivation for doing this is to uncover interesting properties and methods of controlling charged particles on helicoid-shaped nanostructures using magnetic fields.

In this paper, we consider the \emph{additional} effect of an external magnetic field on a charged \emph{spinless} particle on a helicoid. We derive exact expressions for the effective potentials for a charged particle confined on a helicoid for two simple circumferential magnetic field configurations. From these expressions we look at the effect of increasing or decreasing the magnetic field strength on the behavior of the effective potentials; in particular we pay especially close attention to whether the potentials become \emph{attractive} or \emph{repulsive} upon increasing or decreasing the magnetic field strength. We also obtain approximate expressions for the energy levels of the particle for certain values of the angular momentum. What we find is that changing the magnetic field strength alters the behavior of the effective potentials and consequently that of the particle; it is possible to make the potential attractive or repulsive simply by changing the field strength. This points to a way of controlling the average behavior of an ensemble of charged particles on helicoid-shaped nanostructures in the non-interacting limit, a result that will possibly be of much use in areas such as microelectronics and nanochemistry.

However, we consider only a very limited subset of all possible magnetic field configurations. This is because only a small number of field configurations lend themselves to a purely analytic approach. For the majority of configurations the resulting equations of motion are non-separable and require numerical methods to solve. Our focus is on the effective potentials, which govern the quantum dynamics of the charged particles on the helicoid; we will not concern ourselves with the extremely difficult task of deriving exact solutions of the Schrodinger equation.

\section{The effective potential from geometry and magnetic fields}
The Schrodinger equation describing the \emph{surface} wavefunction $\psi_s$ of a charged spinless particle with mass $m$ and charge $Q$ on a two-dimensional curved surface in a magnetic field is given by \cite{gF08}
\begin{widetext}
\begin{equation}
i\hbar\partial_t\psi_s = \frac{1}{2m}\left[-\frac{\hbar^2}{\sqrt{g}}\partial_a(\sqrt{g}g^{ab}\partial_b\psi_s ) + \frac{iQ\hbar}{\sqrt{g}}\partial_a(\sqrt{g}g^{ab}A_b)\psi_s\right]  +  \frac{1}{2m}\left[2iQ\hbar g^{ab}A_a\partial_b\psi_s + Q^2g^{ab}A_aA_b\psi_s   \right] + V_s\psi_s. \label{E:TDSEheli}
\end{equation}
\end{widetext}
Note that Einstein summation notation has been used. In Eq. (\ref{E:TDSEheli}) $\hbar$ is Planck's constant, $\partial_i$ is the partial derivative operator taken with respect to the variable $i$, and $A_j$ denotes the component of the magnetic vector potential $\mathbf{A}$ in the direction of $j$. We allow both $i$ and $j$ are to be equal to $\rho$ and $z$.  If we consider only a magnetic field that is constant in time, then there are no time-dependent terms in the full equation of motion of the wavefunction, and we may therefore work with the time-independent version of Eq. (\ref{E:TDSEheli}):
\begin{widetext}
\begin{equation}
E\chi_s = \frac{1}{2m}\left[-\frac{\hbar^2}{\sqrt{g}}\partial_a(\sqrt{g}g^{ab}\partial_b\chi_s ) + \frac{iQ\hbar}{\sqrt{g}}\partial_a(\sqrt{g}g^{ab}A_b)\chi_s\right]  + \frac{1}{2m}\left[2iQ\hbar g^{ab}A_a\partial_b\chi_s + Q^2g^{ab}A_aA_b\chi_s   \right] + V_s\chi_s \label{E:TISEheli}
\end{equation}
\end{widetext}
where
\begin{equation} \label{E:separatedheli}
\psi_s = \chi_s e^{-iEt/\hbar}
\end{equation}
and $E$ is the energy of the particle. Now we fill in the blanks and put in the expressions for the metric components that enter into the surface equation of motion. These will be specific to the helicoid.

A helicoid can be parametrized by the following set of equations \cite{aG93}:
\begin{eqnarray}
x &=& \rho\cos(\omega z), \label{E:xparaheli} \nonumber \\
y &=& \rho\sin(\omega z), \label{E:yparaheli} \nonumber \\
z &=& z \label{E:zparaheli}
\end{eqnarray}
In these equations $\omega=2\pi S$, where $S$ is the number of complete twists (i.e., $2\pi$-turns) per unit length of the helicoid and $\rho$ is the radial distance from the $z$-axis. For convenience, we use the coordinates $(z, \rho)$ to characterize a point on the helicoid. The infinitesimal line element on the helicoid is given by
\begin{equation} \label{E:lineelementheli}
ds^2 = d\rho^2 + (1 + \omega^2\rho^2)dz^2.
\end{equation}
For ease we let 
\begin{equation}
a(\rho) = \sqrt{1 + \omega^2\rho^2}. 
\end{equation}
The metric components are thus given by
\begin{eqnarray}
g_{\rho \rho} &=& 1, \nonumber \\
g_{zz} &=& 1 + \omega^2\rho^2 = a^2, \nonumber \\
g_{z \rho} &=& 0, \nonumber \\
g_{\rho z} &=& 0, \label{E:metriczheli}
\end{eqnarray}
and the square root of the determinant of the metric is given by
\begin{equation} \label{E:metricdeterminantheli}
\sqrt{g} = \sqrt{1 + \omega^2\rho^2} = a.
\end{equation}
\begin{figure}
\includegraphics[width=0.5 \textwidth]{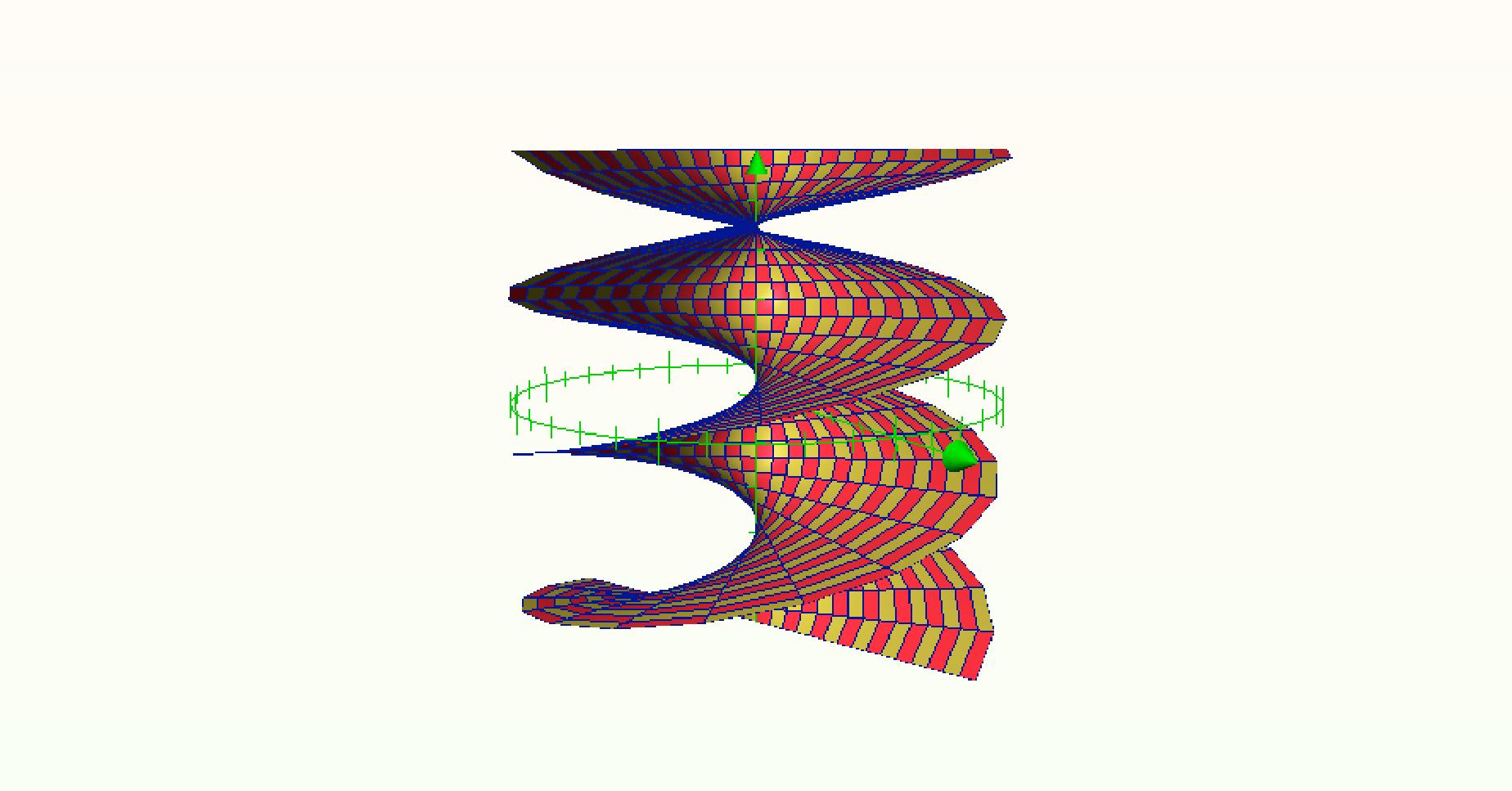}
\caption{A helicoid} \label{helicoidpng} 
\end{figure} 
This gives rise to the following expressions for the principal curvatures:
\begin{eqnarray}
\kappa_1 &=& \frac{\omega}{1+\omega^2\rho^2}, \nonumber \\
\kappa_2 &=& -\frac{\omega}{1+\omega^2\rho^2}. \label{E:curvatureheli}
\end{eqnarray}
As a result, the mean curvature $M$ vanishes:
\begin{equation} \label{E:meancurvatureheli}
M = \frac{1}{2}(\kappa_1 + \kappa_2) = 0.
\end{equation}
This vanishing of $M$ is the reason for calling the helicoid a minimal surface. The Gaussian curvature $K$, on the other hand, is nonvanishing:
\begin{equation} \label{E:gaussiancurvatureheli}
K = \kappa_1\kappa_2 = -\frac{\omega^2}{(1+\omega^2\rho^2)^2}.
\end{equation}
Therefore the curvature-induced potential $V_s$ is given by
\begin{equation} \label{E:curvpotentialheli}
V_s = -\frac{\hbar^2}{2m}\left(M^2 - K \right) =  -\frac{\hbar^2}{2m}\frac{\omega^2}{(1+\omega^2\rho^2)^2}.
\end{equation}

We now turn our attention to the Hamiltonian $H$ of the particle. To clean up our analysis we separate the Hamiltonian into two parts: $H_{em}$, which contains the terms describing the interaction with the magnetic field, and $H_{curv}$, which contains the non-electromagnetic terms---which in our case include the kinetic energy terms and the curvature-induced potential.  The non-electromagnetic part of $H$ is given by
\begin{widetext}
\begin{equation} \label{E:hamiltoniancurv}
H_{curv}\chi_s = \frac{1}{2m}\left[\frac{-\hbar^2}{a} \left(  \partial_z( \frac{1}{a}\partial_z\chi_s ) + \partial_{\rho}(a\partial_{\rho}\chi_s )             \right)  \right] + V_s\chi_s.
\end{equation}
\end{widetext}
Since the wavefunction has to be normalized with respect to the infinitesimal area $d\rho dz$---and not with respect to $ad\rho dz$, as Eq. (\ref{E:metriczheli}) would suggest---we make the substitution $\chi_s \to \frac{1}{\sqrt{a}}\chi_s$ in Eq. (\ref{E:hamiltoniancurv}), which would only affect terms involving derivatives with respect to $\rho$. After lengthy algebra, we arrive at this expression for $H_{curv}$:
\begin{widetext}
\begin{equation} \label{E:hamiltoniancurvnorm}
H_{curv}\chi_s = -\frac{\hbar^2}{2m} \left[\frac{\partial^2\chi_s}{\partial\rho^2}  + \frac{1}{1+\omega^2\rho^2} \frac{\partial^2\chi_s}{\partial z^2}  + \frac{\omega^2}{2(1+\omega^2\rho^2)^2} \left(1 + \frac{\omega^2\rho^2}{2} \right)\chi_s \right].
\end{equation}
\end{widetext}
Note that this expression is simply the \emph{full} Hamiltonian for an uncharged particle on a helicoid, as derived in the study by Atanasov \emph{et al.} \cite{vA09}.  

$H_{em}$ meanwhile is given by 
\begin{widetext}
\begin{equation}
H_{em}\chi_s = \frac{iQ\hbar}{2m} \left[ \frac{1}{1+\omega^2\rho^2} \frac{\partial A_z}{\partial z}  + \frac{\partial A_{\rho}}{\partial \rho}  + \frac{\omega^2\rho}{1+\omega^2\rho^2}A_{\rho} \right]\chi_s + \frac{2iQ\hbar}{2m}\left[\frac{A_z}{1+\omega^2\rho^2}\frac{\partial\chi_s}{\partial z} + A_{\rho}\frac{\partial\chi_s}{\partial\rho}    \right] + \frac{Q^2}{2m}\left( \frac{A_z^2}{1+\omega^2\rho^2} + A_{\rho}^2  \right)\chi_s. \label{E:hamiltonianem}
\end{equation}
\end{widetext}
If we work with magnetic vector potentials in which $A_{\rho} = 0$, then $H_{em}$ \emph{remains} the same even after the wavefunction $\chi_s$ has been adjusted with the $1/\sqrt{a}$ factor, since the term involving the derivative of the wavefunction with respect to $\rho$ vanishes.

We now have the full Hamiltonian, and consequently the time-independent Schrodinger equation $E\chi_s = H\chi_s$, for a charged spinless particle confined on a helicoid. The Schrodinger equation we have derived is a partial differential equation in two variables and is extremely difficult to solve in general. However, for some field configurations, separation of variables can be done, and we are left with a simpler ordinary differential equation. We now consider specific field configurations.

\section{Behavior of the effective potential}
\subsection{Constant circumferential magnetic field}

A magnetic field that is circumferential and of constant magnitude everywhere can be described by the vector equation 
\begin{equation}
\mathbf{B} = B\mathbf{\hat{\phi}}, 
\end{equation}
where $B$ is a constant. This is not a very realistic field configuration, as it follows from Maxwell's equations that no \emph{physically realizable} current density $\mathbf{j}$ will produce this magnetic field; such a current density would have to take on an infinite value at the center of the wire. Nevertheless, this field configuration is useful for two reasons: 1) It provides the simplest test case for the formalism that we have developed for the charged particle on a helicoid, and the calculation is instructive and reveals a lot about the interplay between the geometric potential and the magnetic field; and 2) it is a reasonable approximation when one considers physically realizable circumferential fields that are sufficiently far from sources, in which case the magnetic field does $\emph{not}$ change much over the relevant length scale. An example of this would be the magnetic field produced by a current-carrying wire (which we will consider in the next section); when we consider portions of the helicoid sufficiently far from the current, the field does not change appreciably with changes in the distance, and therefore in this regime it can be reasonably approximated by a constant circumferential field.

\subsubsection{The Schrodinger equation and the effective potentials}

We choose the most convenient vector potential $\mathbf{A}$ that gives $\mathbf{B} = B\mathbf{\hat{\phi}}$. The components of $\mathbf{A}$ are given by
\begin{equation} \label{E:avectorpotentialheli}
A_z = -B\rho, \quad A_{\rho} = 0.
\end{equation}
Substituting this into Eq. (\ref{E:hamiltonianem}) and combining with Eq. (\ref{E:hamiltoniancurvnorm}), we have the following expression for the full Schrodinger equation:
\begin{widetext}
\begin{equation}
E\chi_s = -\frac{\hbar}{2m}\left(\frac{\partial^2\chi_s}{\partial\rho^2} + \frac{1}{1+\omega^2\rho^2} \frac{\partial^2\chi_s}{\partial z^2}\right) - \frac{\hbar^2}{2m}\left[\frac{\omega^2}{2(1+\omega^2\rho^2)^2} \left(1 + \frac{\omega^2\rho^2}{2}\right)   \right] \chi_s - \frac{2iQ\hbar B\rho}{2m(1+\omega^2\rho^2)} \frac{\partial\chi_s}{\partial z} + \frac{Q^2B^2\rho^2}{2m(1+\omega^2\rho^2)} \chi_s. \label{E:hamiltonsub}
\end{equation}
\end{widetext}
The absence of $z$-dependent terms in Eq. (\ref{E:hamiltonsub}) gives us license to separate $\chi_s$ in the following manner \cite{LL77}:
\begin{equation} \label{E:splitwaveheli}
\chi_s = e^{ikz}\gamma(\rho).
\end{equation}
The wavevector $k$ is related to the $z$-component of the momentum $p_z$ by the relation $k = p_z/\hbar$. Since $\phi = \omega z$ for a helicoid, it can be shown that $p_z = \omega L_{\phi}$, where $L_{\phi}$ is the angular momentum about the $z$-axis. From the periodicity of the system, we require that $e^{ikz}$ be periodic with period $1/S$; this implies that $k = 2\pi S l = \omega l$, where $l \in \mathbb{Z}$.

Putting Eq. (\ref{E:splitwaveheli}) into Eq. (\ref{E:hamiltonsub}), we have
\begin{widetext}
\begin{equation}
 E\gamma = -\frac{\hbar}{2m}\frac{d^2\gamma}{d\rho^2} - \frac{\hbar}{2m}\left(\frac{\omega^2}{2(1+\omega^2\rho^2)}\right)\left(1 + \frac{\omega^2\rho^2}{2}   \right)\gamma + \frac{\hbar^2 l^2\omega^2}{2m(1+\omega^2\rho^2)}\gamma + \frac{2Q\hbar B\rho l\omega}{2m(1+\omega^2\rho^2)}\gamma + \frac{Q^2 B^2 \rho^2}{2m(1+\omega^2\rho^2)}\gamma. \label{E:eqmotionheli}
\end{equation}
\end{widetext}
After simplifying, $V$ can be shown to be given by the following expression:
\begin{widetext}
\begin{equation}
V(\rho) = \left(-\frac{\hbar^2\omega^2}{8m} \right)\left\{\frac{1}{(1+\omega^2\rho^2)^2} + \frac{1}{1+\omega^2\rho^2}\left[1 - 4l^2 + \frac{4Q^2B^2}{\hbar^2\omega^4} \right]\right\}  + \left(-\frac{\hbar^2\omega^2}{8m} \right)\left\{\frac{\rho}{1+\omega^2\rho^2}\left(-\frac{8QBl}{\hbar\omega} \right)  -\frac{4Q^2B^2}{\hbar^2\omega^4}  \right\}. \label{E:potentialheli}
\end{equation}
\end{widetext}
To simplify the form of Eq. (\ref{E:potentialheli}), we introduce the following parameters:
\begin{eqnarray}
\tau &=& \frac{8QB}{\hbar}, \\
\quad \xi &=& \frac{\hbar^2}{8m}.
\end{eqnarray}
We add that $\tau$ is a convenient proxy for the applied magnetic field strength. Noting that $\tau^2 = 64Q^2B^2/\hbar^2$, Eq. (\ref{E:potentialheli}) becomes, in terms of $\tau$ and $\xi$,
\begin{widetext}
\begin{equation} \label{E:potentialhelisimple}
V(\rho) = -\xi\omega^2\left\{\frac{1}{(1+\omega^2\rho^2)^2} + \frac{1}{1+\omega^2\rho^2}\left[1 - 4l^2 + \frac{\tau^2}{16\omega^4} \right] - \frac{\rho}{1+\omega^2\rho^2}\left(\frac{\tau l}{\omega} \right) - \frac{\tau^2}{16\omega^4}  \right\}.
\end{equation}
\end{widetext}
Clearly the behavior of $V$ depends on the applied magnetic field strength and on the angular momentum mode of the particle. We note that when $\tau = 0$, we regain the effective potential of a particle on a helicoid $\emph{without}$ an external applied field, as found in a previous study \cite{vA09}.  We plot the potentials for a particle on a helicoid for both positive and negative $\mathbf{B}$ in Figure 5.2 and Figure 5.3, respectively. 

\begin{figure*}  \label{F:graph_heli1}
\includegraphics[width=1 \textwidth]{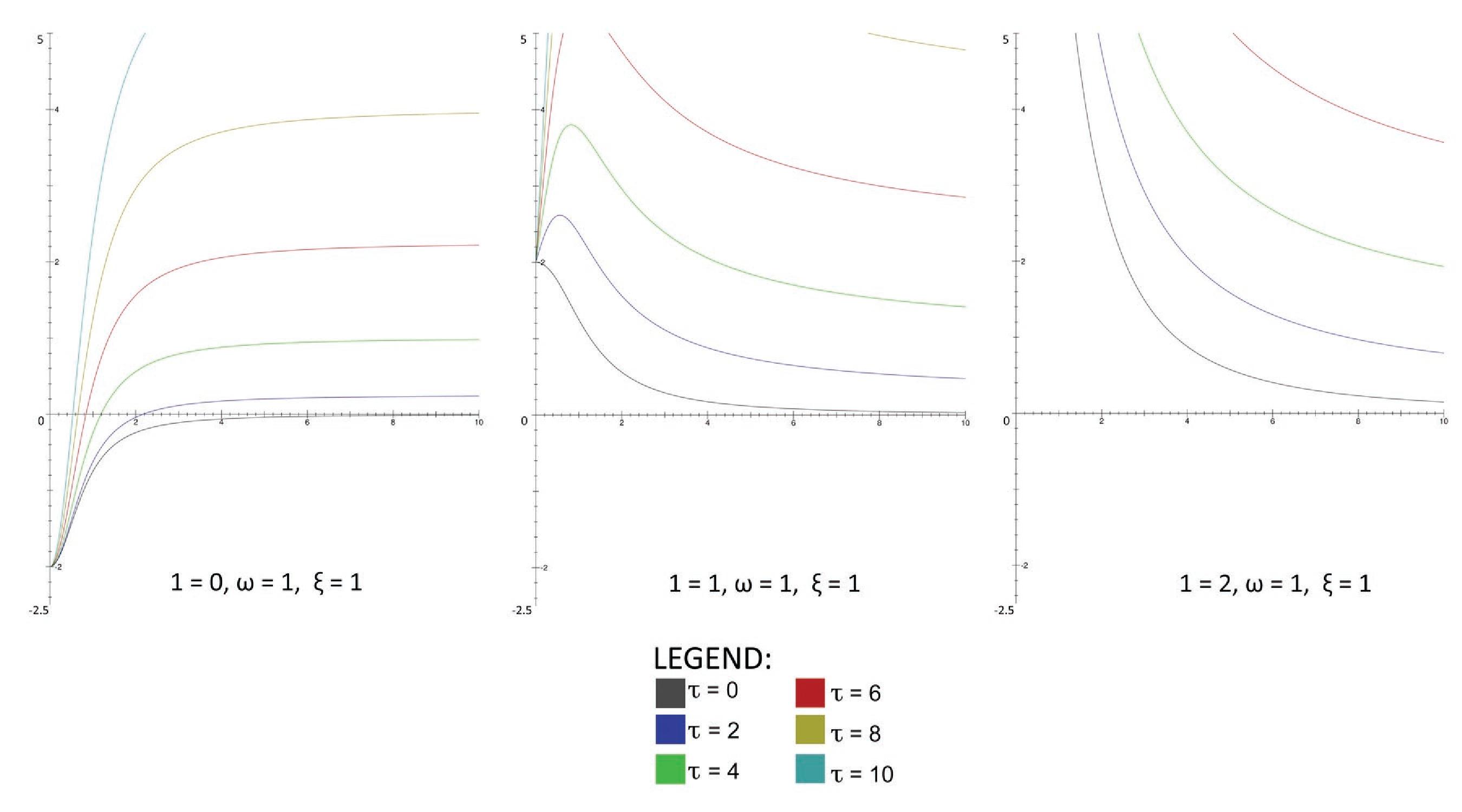}
\caption{The effective potential $V$ (\emph{y}-axis) with respect to $\rho$ (\emph{x}-axis) for a particle on a helicoid in a constant magnetic field, for various angular momentum modes and $QB>0$ (that is, $\emph{positive}$ magnetic field for $Q > 0$, or $\emph{negative}$ magnetic field for $Q < 0$).} \label{helicoidconst+} 
\end{figure*} 

\begin{figure*} \label{F:graph_heli2}
\centering
\includegraphics[width=1 \textwidth]{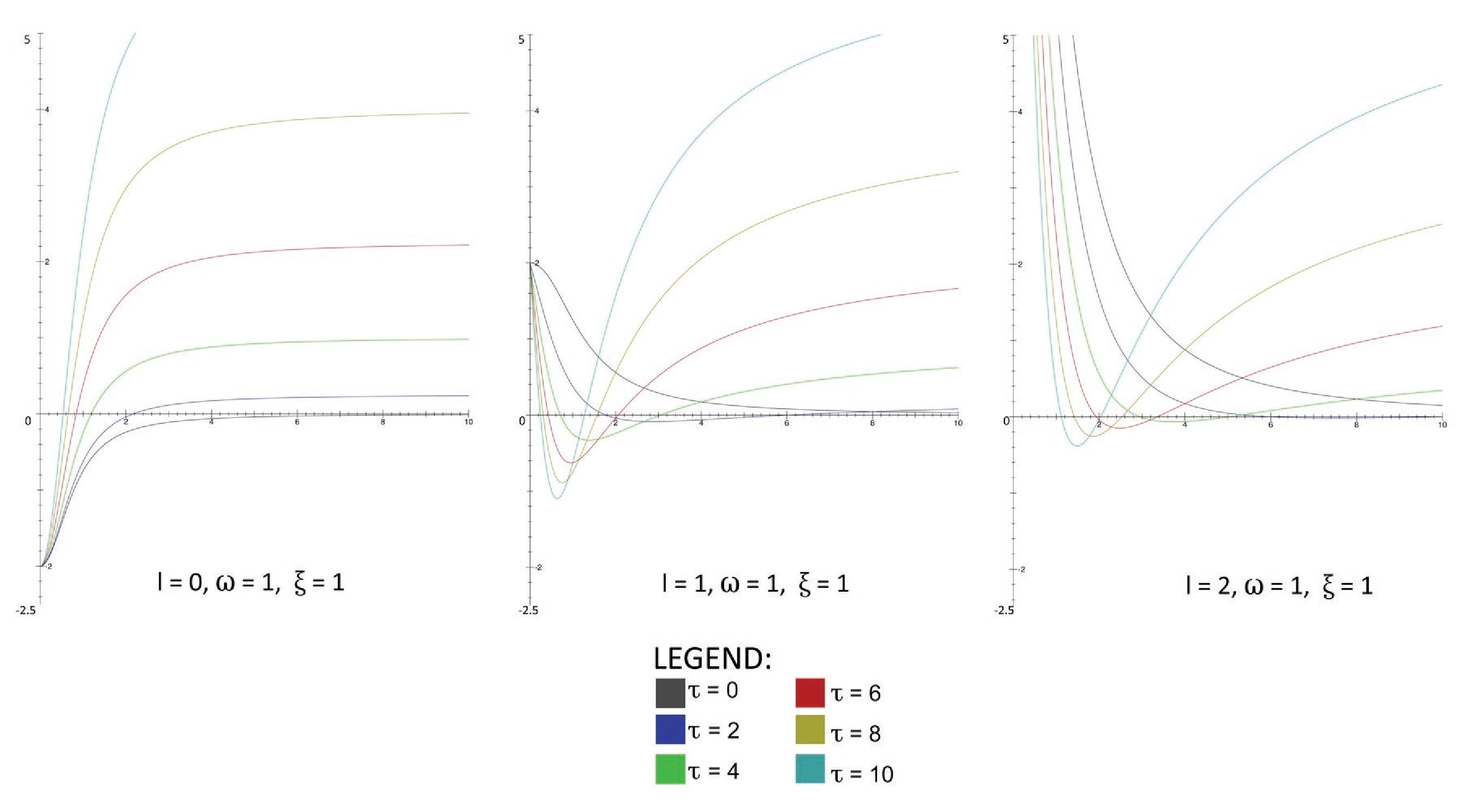}
\caption{The effective potential $V$ (\emph{y}-axis) with respect to $\rho$ (\emph{x}-axis) for a particle on a helicoid in a constant magnetic field, for various angular momentum modes and $QB<0$ (that is, $\emph{negative}$ magnetic field for $Q > 0$, or $\emph{positive}$ magnetic field for $Q < 0$).} \label{helicoidconst-} 
\end{figure*} 

\subsubsection{Behavior of the effective potential}

Let us consider the case $l = 0$. The effective potentials are identical for both positive and negative magnetic field strengths. For the zero-field case, the potential is attractive towards the origin---in fact, there is a shallow potential well near the origin. When the magnetic field strength is increased or decreased, the potential well near the origin becomes deeper, such that for extremely strong fields, the particle is practically confined at the origin. 

The case $l = 1$ is more interesting. When the magnetic field is zero, the effective potential is repulsive from a certain distance from the origin onwards. If we have $QB > 0$ (that is, either $Q>0$ and $\mathbf{B}$ is in the direction of $\hat{\phi}$, \emph{or}, equivalently, $Q<0$ and $\mathbf{B}$ is in the opposite direction) \emph{and} the field magnitude $B$ is increased, the potential remains repulsive. The effect thus is that a charged particle is much less likely to be found near the origin. However, different behavior occurs when we have $QB<0$ and then increase the magnitude of $B$. For a sufficiently strong magnetic field with this orientation, the effective potential becomes attractive near the origin, and potential wells form. These mean that charged particles are much more likely to be found near these areas, where the effective potential is a minimum. As the magnetic field is increased in magnitude, the potential wells become deeper; for very strong magnetic fields, the particle becomes localized near (but not $\emph{at}$) the origin. Also, the distance at which the potential is a minimum becomes less when the magnetic field strength is increased. When $l = 2$, the results are for the most part qualitatively similar to that when $l = 1$---only this time, when $QB>0$, the potential is repulsive for nearly all $\rho$; and when $QB<0$, the potential minima are shifted some distance outward.

For a large ensemble of charge carriers in the non-interacting limit distributed over a helicoid, the $QB<0$ scenario would cause most of these charge carriers to be located along a narrow strip near the radius $\rho_{min}$ where the potential is a minimum. This can be realized by placing, say, negatively charged carriers such as electrons on a helicoid and applying a magnetic field $\mathbf{B}$ in the $-\hat{\phi}$ direction. If instead we have $QB>0$---which can be realized by reversing the direction of the magnetic field in the previous example---most of these charge carriers would be located far from the origin, near the outer edge of the helicoid-shaped strip, thanks to the repulsive nature of the effective potential. A possible application of this result would be the use of a magnetic field to control the flow of electric current through a nanostructure; by adjusting the applied magnetic field, one can either make the current stay on a strip near the origin, or make it localized primarily near the outer edge of the helicoid. 

\subsubsection{Energy levels}

In this subsection we perform a rudimentary calculation of the approximate energy levels of the charged particle on a helicoid in a constant circumferential magnetic field.  We do this to see the dependence of the energies on the applied magnetic field strength, and this description should be applicable in general for potentials with a minimum. We restrict our analysis to the case $l = 0$, since for higher values of the angular momentum solving for the minima of the effective potentials entails solving a quartic equation. In those cases it is far more worthwhile to attack the problem using numerical methods instead. 

For $l = 0$, Eq. (\ref{E:potentialhelisimple}) becomes
\begin{widetext}
\begin{equation} \label{E:potential_0}
V(\rho) = -\xi\omega^2\left\{\frac{1}{(1+\omega^2\rho^2)^2} + \frac{1}{1+\omega^2\rho^2}\left[1 + \frac{\tau^2}{16\omega^4} \right] - \frac{\tau^2}{16\omega^4}  \right\}.
\end{equation}
\end{widetext}

We expand Eq. (\ref{E:potential_0}) as a Taylor series about its minimum. Evidently it has a minimum at $\rho = 0$, as can be seen in Figs. (5.2) and (5.3). We consider only small deviations from $\rho = 0$ so that we can ignore third- and higher-order terms in the Taylor expansion. The effective potential near $\rho = 0$ is thus given by

\begin{equation} \label{E:potentialapprox}
V(\rho) \approx -2\xi\omega^2 + \xi\omega^4\left(3 + \frac{\tau^2}{16\omega^4}\right)\rho^2.
\end{equation}

Replacing the exact potential with the approximate one, we have the following expression for the Schrodinger equation:

\begin{equation} \label{E:schrodingersimple}
-\frac{\hbar^2}{2m}\frac{d^2\gamma}{d\rho^2} + \xi\omega^4\left(3 + \frac{\tau^2}{16\omega^4}\rho^2\right)\gamma = \left(E + 2\xi\omega^2\right)\gamma.
\end{equation}

Eq. (\ref{E:schrodingersimple}) is similar in form to the Schrodinger equation for a particle in a harmonic oscillator potential $\frac{1}{2} m\omega^2_0\rho^2$ (with characteristic frequency $\omega_0$), provided that
\begin{equation} \label{characteristicfrequency}
\omega_0 = \sqrt{\frac{2\xi\omega^4}{m}\left(3 + \frac{\tau^2}{16\omega^4}\right)}.
\end{equation}
The energy levels for this system are thus similar to that of a particle in a harmonic oscillator potential. We have
\begin{equation} \label{energylevels}
E_n = \hbar\sqrt{\frac{2\xi\omega^4}{m}\left(3+\frac{\tau^2}{16\omega^4}\right)}\left(n + \frac{1}{2}\right) - 2\xi\omega^2,
\end{equation}
or, expressing this in terms of the original variables $Q$ and $B$,
\begin{equation} \label{origenergylevels}
E_n = \frac{\hbar^2\omega^2}{2m}\sqrt{3 + \frac{4Q^2B^2}{\hbar^2\omega^4}}\left(n + \frac{1}{2}\right) - \frac{\hbar^2\omega^2}{4m}.
\end{equation}
The term $-\hbar^2\omega^2/4m$ in the energies is due to the constant term $-\tau^2/16\omega^4$ in the effective potential given in Eq. (\ref{E:potentialhelisimple}), which adds a time-dependent phase factor in the wavefunction but has no effect on the expectation values of any dynamical variable.

Eq. (\ref{energylevels}) gives the approximate energy levels for a charged particle with angular momentum corresponding to $l = 0$. Clearly, increasing $\mathbf{B}$ increases the energy $E_n$ for all $n$; for large $B$ the energy scales as $E \sim B$.

\subsection{Magnetic field from a current-carrying wire}
Perhaps the most physically relevant and experimentally realizable field configuration is that of a magnetic field produced by a current-carrying wire. We place the wire in such a way that it is in the middle of the helicoid; that is, the wire coincides with the $z$-axis. If a current $I$ in the $z$-direction flows through the wire, then it will create a circumferential magnetic field, given by
\begin{equation} \label{E:magneticfieldwire}
\mathbf{B} = \frac{\mu_0 I}{2\pi\rho}\hat{\phi}.
\end{equation}
For convenience we let 
\begin{equation}
\delta = \frac{\mu_0}{2\pi}I. 
\end{equation}
The components of the vector potential $\mathbf{A}$ that gives rise to this magnetic field configuration are given by
\begin{eqnarray} 
A_z &=& -\delta\ln\rho, \nonumber \\
A_{\rho} &=& 0. \label{E:vectorpotentialwire}
\end{eqnarray}
Repeating the same steps done in the previous section, we have the following Schrodinger equation for the surface wavefunction $\chi_s$:
 \begin{widetext}
 \begin{equation}
E\chi_s = -\frac{\hbar}{2m}\left(\frac{\partial^2\chi_s}{\partial\rho^2} + \frac{1}{1+\omega^2\rho^2} \frac{\partial^2\chi_s}{\partial z^2}\right) - \frac{\hbar^2}{2m}\left[\frac{\omega^2}{2(1+\omega^2\rho^2)^2} \left(1 + \frac{\omega^2\rho^2}{2}\right)   \right] \chi_s  - \frac{2iQ\hbar\delta\ln\rho}{2m(1+\omega^2\rho^2)} \frac{\partial\chi_s}{\partial z} + \frac{Q^2\delta^2(\ln\rho)^2}{2m(1+\omega^2\rho^2)} \chi_s. \label{E:schrodingereqwire}
\end{equation}
\end{widetext}
 
 As before, we exploit  the periodicity of the system and the absence of $z$-dependent terms in Eq. (\ref{E:schrodingereqwire}). We introduce the ansatz  $\chi_s = e^{ikz}\gamma(\rho)$ and substitute this into the Schrodinger equation. After a lot of algebra, we have
 \begin{widetext}
 \begin{equation} \label{E:schrodingereqfinal}
-\frac{\hbar}{2m}\frac{d^2\gamma}{d\rho^2} - \frac{\hbar^2\omega^2}{8m}\left[\frac{1}{(1+\omega^2\rho^2)^2} + \frac{1 - 4l^2}{1+\omega^2\rho^2}\right]\gamma + \frac{2Q\delta\hbar\omega l \ln(\rho) }{m(1+\omega^2\rho^2)}\gamma + \frac{Q^2\delta^2(\ln\rho)^2}{2m(1+\omega^2\rho^2)}\gamma = E\gamma.
\end{equation}
\end{widetext}
All the terms in the left hand side except the first represent the effective potential $V$ of the system. If we introduce the dimensionless variables
\begin{eqnarray}
\zeta = 8Q\delta/\hbar, \nonumber \\
\lambda ' = \hbar^2/8m, 
\end{eqnarray}
it can be shown that $V$ can be expressed in the simpler form
\begin{equation} \label{E:potentialwire}
V = -\lambda ' \omega^2\left[\frac{1}{(1+\omega^2\rho^2)^2} + \frac{1 - 4l^2 - \frac{\zeta l \ln\rho}{\omega} - \frac{\zeta^2(\ln\rho)^2}{16\omega^2}   }{1+\omega^2\rho^2} \right].
\end{equation}
In Figure 5.4 we plot the effective potential of the particle for a variety of angular momentum modes and magnitudes of the source current (as represented by the variable $\delta$).

\begin{figure*} \label{F:graph_3}
\includegraphics[width=1 \textwidth]{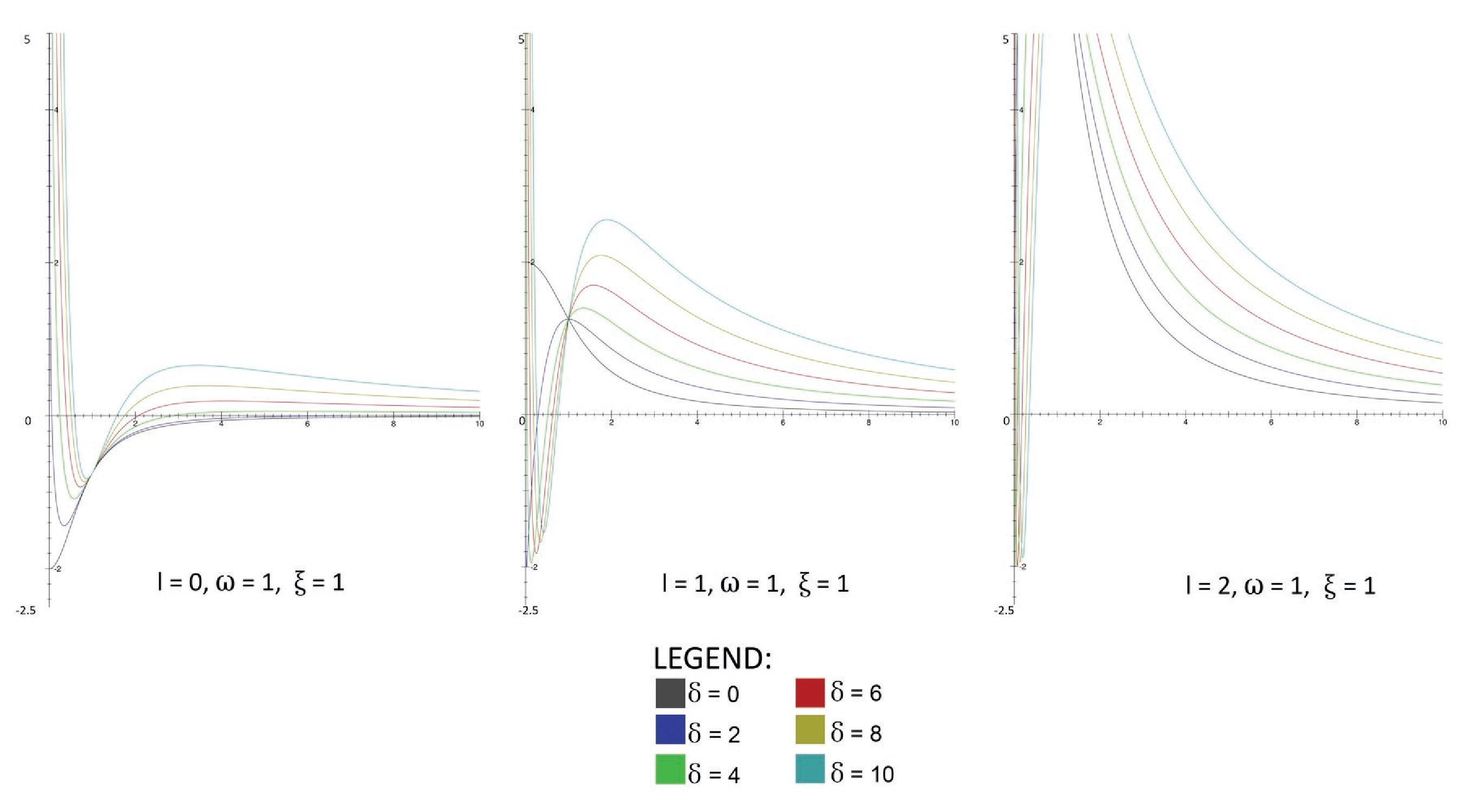}
\caption{The effective potential $V$ (\emph{y}-axis) with respect to $\rho$ (\emph{x}-axis) for a particle on a helicoid in the magnetic field of a current-carrying wire, for variable angular momentum modes and current $I > 0$ (for $Q > 0$), or, \emph{equivalently,} $I<0$ (for $Q<0$). } \label{helicoidwire} 
\end{figure*} 

When $l = 0$, the zero-field potential is attractive throughout, and is a minimum at the origin. Let us consider the case where $\zeta > 0$ (that is, either $I>0$ and $Q>0$, or $I<0$ and $Q<0$. We would then see that radius at which the effective potential is a minimum is shifted outward. For a wide range of current magnitudes, it is found that potential wells are formed near the origin. When the magnitude of the source current is increased, the radius at which the potential is a minimum is moved farther away from the center of the helicoid. For a large ensemble of charge carriers, the application of a magnetic field from a line current source would mean that the charge carriers will be clustered along a strip where the effective potential is a minimum.

The effective potentials for the higher angular momentum modes $l = 1$ and $l = 2$ display interesting behavior as well. The effective potential when there is no magnetic field is repulsive throughout the helicoid---which means that the particle will be more likely to be located on the outer rim of the helicoid. The addition of a magnetic field from a current source however causes the potential to be attractive near the origin and repulsive far from it. In these higher angular momentum modes the potential wells are much deeper and narrower than in the $l = 0$ case. The distances at which the effective potentials are minima for these angular momentum modes are also much smaller than in the $l = 0$ case. These properties are especially apparent in the $l = 2$ case, where the potential wells near the origin are much deeper and narrower than those corresponding to the $l = 0$ and $l = 1$ modes. 

This implies that for higher angular momentum modes, the particle is effectively localized along a thin strip of the helicoid corresponding to where the potential is a minimum. If we consider a large number of charge carriers, this would mean that we will see most of the charged particles to be clustered around a narrow strip of the helicoid. For higher angular momentum modes the width of this strip is much smaller than in the $l = 0$ mode, and the strip will be found much closer to the center of the helicoid.

This system can be realized experimentally by creating a helical nanostructure, putting charge carriers on the nanostructure, and placing a tiny current-carrying wire through the center of the helicoid. Since the resulting potential wells are deep, the application and modulation of a magnetic field provides a method for the control of current in this helicoid-shaped nanostructure.

\section{Conclusions}

We have considered the quantum mechanics of a charged particle on a helicoid in an external magnetic field. We derive the equations of motion for the wavefunction of a charged particle confined for two simple circumferential magnetic field configurations, and from these obtain expressions for the effective potentials that govern the behavior of the particle. We plot these potentials and examine their behavior for different values of angular momentum and the applied magnetic field strength. We were also able to derive approximate expressions for its energy levels for certain values of the angular momentum, to see the dependence of the energies on the applied magnetic field strength.

These effective potentials are dependent on both the surface curvature and the strength of the applied magnetic field. By changing the strength of the magnetic field, we observe that it is possible to change the qualitative behavior of the potential, which can range from repulsive to attractive for varying magnetic field strength and direction.  Meanwhile, the energy levels of the charged particle on a helicoid are found to be similar in form to that of a particle in a harmonic oscillator potential, for potentials which have a minimum. For the case of a large ensemble of charge carriers on a helicoid-shaped nanostructure in the non-interacting limit, the application of an applied magnetic field would lead, for a range of magnetic field strengths, to the clustering of charge carriers on a strip near areas where the effective potential is a minimum. By increasing or decreasing the magnetic field strength, we find that it is possible to control where the strip is located and how thick this strip is. 

There are many avenues for further research in this topic. A lot of electric and magnetic field configurations which do not yield separable equations of motion require a closer examination, ideally with numerical methods. Also, the effect of the curvature on the nonrelativistic quantum dynamics of charged particles has not yet been studied for many surfaces. More complex effects arising from the combination of the geometric potential and the electromagnetic field may result.

\begin{acknowledgements}
M.A.S. is very grateful to Germelino Abito for his support and advice in the initial stages of this project, and to Raphael Guerrero and Job Nable for very helpful discussions.
\end{acknowledgements}

\end{document}